\documentclass[preprint,5p,times,twocolumn]{elsarticle}

\usepackage[table,xcdraw]{xcolor}
\usepackage{amsmath}
\usepackage{listings}
\usepackage{booktabs}
\usepackage{makecell}
\usepackage[flushleft]{threeparttable}
\usepackage{url}

\newcolumntype{P}[1]{>{\centering\arraybackslash}p{#1}}

\journal{Journal of Systems and Software}

\begin{document}
\begin{frontmatter}

\title{Automatic Extraction of Security-Rich Dataflow Diagrams \\for Microservice Applications written in Java}

\author{Simon Schneider}
\ead{simon.schneider@tuhh.de}

\author{Riccardo Scandariato}
\ead{riccardo.scandariato@tuhh.de}
\affiliation{organization={Institute of Software Security, Technical University Hamburg},%Department and Organization
            city={Hamburg},
            country={Germany}}

\begin{abstract}
Dataflow diagrams (DFDs) are a valuable asset for securing applications, as they are the starting point for many security assessment techniques. 
Their creation, however, is often done manually, which is time-consuming and introduces problems concerning their correctness.
Furthermore, as applications are continuously extended and modified in CI/CD pipelines, the DFDs need to be kept in sync, which is also challenging. 
In this paper, we present a novel, tool-supported technique to automatically extract DFDs from the implementation code of microservices. 
The technique parses source code and configuration files in search for keywords that are used as evidence for the model extraction.
Our approach uses a novel technique that iteratively detects new keywords, thereby \textit{snowballing} through an application's codebase.
Coupled with other detection techniques, it produces a fully-fledged DFD enriched with security-relevant annotations.
The extracted DFDs further provide full traceability between model items and code snippets.
We evaluate our approach and the accompanying prototype for applications written in Java on a manually curated dataset of 17 open-source applications.
In our testing set of applications, we observe an overall precision of 93\% and recall of 85\%.
\end{abstract}


\end{frontmatter}

\section{Introduction}
\label{sec:introduction}

\noindent The microservice architecture has seen increasing adoption over the last years as it addresses some problems that modern, often cloud-based systems face when following a monolithic approach~\cite{Dragoni16}. Separating an application into \textit{microservices} (or simply \textit{services}) such that each service realizes a single business functionality and is implemented as an independent system with dedicated resources is advantageous for its scalability, deployment, development, and maintenance~\cite{Chen17,Jamshidi18}.
However, this also introduces additional security challenges, which lie mainly in securing the extended attack surface, in the inter-service communication channels needed to achieve the business functionality, and in establishing trust between services~\cite{Hannousse21,Li19,Pereira-Vale21,Yarygina18}.

Many of these challenges can be tackled by analysing the architectural model of the microservice applications (in short \textit{microservices}) and by identifying the possible security weaknesses.
However, a suitable model needs to be available.
In this paper, we focus on dataflow diagrams (DFDs) for representing microservice applications, since they are the starting point for many security assessment techniques, like threat analysis~\cite{Sion18, Hernan06, Microsoft16, Torr05} and automated model analysis \cite{Abi-Antoun07, Abi-Antoun10, Berger16, Tuma19}.
In fact, DFDs have been described as \textit{threat model diagrams} because of their frequent use for this purpose~\cite{Shostack14}.
Additionally, since the architecture maps well to DFDs, they are also used specifically for microservices~\cite{Chen17, Stojanovic20, Li19c}.
DFDs depict the system components as nodes and the connections between them as directed edges. 
However, the DFD notation is rather minimalist and needs to be augmented with annotations that can give further details on components and connections, for example about the deployed security mechanisms or other properties that are useful for analysis. 

Although they can have positive impact on application security, DFDs are often seen by developers as being tedious and time-consuming to create. 
Bernsmed et. al \cite{Bernsmed21} triangulated four studies to derive benefits and obstacles of using DFDs in companies. 
The results show that developers, on average, do not like creating them and struggle with details, for example finding the correct granularity or knowing what to include in the diagrams. 
The microservice architecture brings further obstacles, since (i) applications can be comprised of large numbers of  services that are loosely coupled by design, and (ii) their deployment is fluid, with service instances being deployed or taken down regularly. 
This situation results in an increased cognitive load for the developers or security analysts that are tasked with creating DFDs of microservice systems.
Furthermore, the microservice approach is often used in the context of CI/CD pipelines, where the application is modified continuously, hence creating a potential disconnection between (manually created) models and (fast changing) code over time.

\textbf{Contributions.} In this paper, we present a novel approach that extracts security-rich DFDs from microservices' source code written in Java.
To the best of our knowledge, this is the first paper focusing on DFDs and microservices for security.
As we show in Section \ref{sec:related_work}, some approaches exist that are able to extract the basic architecture of microservices.
We decide against extending one of these in favor of developing an approach of our own, since the model items \emph{beyond} the basic structure are the challenging part in our work and, thus, we opt for having a coherent approach and implementation.
We choose Java as the target language for our prototype, since a search on GitHub for terms such as `microservice' retrieves the highest number of results in Java, indicating it to be the most used language for this purpose at the moment.
Further, we focus on applications using the Spring framework \footnote{https://spring.io -- In this paper, we refer to the different Spring projects (e.g. Boot, MVC, Cloud, Security) collectively as `Spring framework'}, as it is by far the most used framework for developing microservices in Java
according to a recent report which states that 74\% of developers that work on microservices use Spring for their work \cite{JRebel22}.

Our model extraction technique is based on textual analysis (i.e., code as text) where the source code and the configuration files are statically parsed in search for keywords that, once found, are used as evidence for the creation of corresponding model items.
The novel technique that we employ iteratively detects new keywords to connect known commands with developer-chosen identifiers, thereby \textit{snowballing} through an application's codebase.
Combined with a direct keyword search and parsing of configuration-, build-, and IaC-files (Infrastructure as Code), the outcome of our approach is a fully-fledged DFD with an extensive set of presented security features.
The approach is able to extract the structure of the application, including (i) the services developed to implement the business logic, (ii) the support services that are deployed in the infrastructure (databases, registry, API gateways and so on), as well as (iii) the information flows among these services.
Additionally, we locate the security features (e.g., use of encryption, authentication, load balancers, and so on) and other system properties in the codebase~\footnote{In this paper, we jointly refer to both Java and configuration files as `code'.} and, accordingly, enrich the DFD model with security annotations and other information that might be relevant for an application assessment from a security standpoint.
In this respect, the \textbf{challenge} that we faced in developing the approach lies in the detection of (i) inter-service communication, which is often implicit and requires the analysis of numerous components across the codebase to grasp all properties of a single connection, and (ii) the identification of many security and availability mechanisms, which are implicitly deployed within other techniques.
For example, when detecting information flows via message brokers, the sending and receiving service and the message broker service are all possible places where configurations of queues/exchanges, etc. can exist, which influence the security properties of a connection.

\textbf{Benefits.} The benefits of our technique are twofold.
First, it is lightweight and runs fast (on average under five seconds).
Therefore, it can be easily integrated in a CI/CD development environment.
Second, it provides explainable results.
In fact, our technique can produce complete traceability information between the generated model and the codebase.
That is, it is possible to view the code snippets that have been used as evidence to create a model item (e.g. an information flow, or a security annotation).
This information is meant to assist both security assessors who need to perform a security analysis of the whole application, as well as developers who want to ascertain the architectural (and security) impact of their commits.

\textbf{Research questions.} This work addresses the following (feasibility related) research questions concerning the extraction of dataflow diagrams from microservice applications written in Java with the technique summarized above:\\

\noindent\fbox{
    \parbox{0.97\linewidth}{
    \begin{description}
        \vspace{-2mm}
        \item [RQ1] What is the precision and recall concerning the extraction of all model items?
        \vspace{-1mm}
        \item [RQ2] What is the precision and recall concerning the extraction of just the DFD core model items (services, databases, external entities, and information flows)?
        \vspace{-1mm}
        \item [RQ3] What is the precision and recall for the extraction of just the annotations that are relevant for security?
        \vspace{-1mm}
        \item [RQ4] What is the execution time for the full extraction?
        \vspace{-3mm}
    \end{description}
    }
}
\vspace{2mm}

\textbf{Validation.} In order to answer the above questions, we implemented a prototype (called \emph{Code2DFD} and released to the open domain \cite{Code2DFD}) and used it in a validation experiment on a pool of open-source applications. 
This was a challenge because no suitable dataset could be found in the public domain.
While there is an abundance of suitable open-source projects, no DFDs could be found in literature where the model elements are traced to the code.
Therefore, in previous work, we created a dataset from scratch, which could be used as ground truth for our experiments \cite{Dataset}.
The evaluation is performed by applying the prototype to the manually curated dataset of 17 microservice applications that have been inspected by four researchers~\footnote{The dataset is available under \textit{https://tuhh-softsec.github.io/microSecEnD/} or via its DOI~\cite{Dataset}}.
For each application, the dataset provides the DFD and traceability information.

\textbf{Paper structure.} The rest of this paper is organized as follows. 
In Section~\ref{sec:mapping} we introduce the DFD items covered by our approach. 
In Section~\ref{sec:approach} we present our novel approach in detail and Section \ref{sec:prototype} describes the prototype we built.
Section~\ref{sec:methodology} describes the evaluation methodology we followed and Section~\ref{sec:results} provides the evaluation results, which are further discussed in Section~\ref{sec:discussion}.
In Section~\ref{sec:comparison}, we compare the performance of our prototype against two related approaches.
Section~\ref{sec:related_work} presents the related work and Section~\ref{sec:conclusion} concludes this paper.

\section{Mapping Microservice Code to a DFD}
\label{sec:mapping}

\begin{figure*}
    \centering
    \caption{Example of extracted model items. Excerpt of the DFD for Piggy Metrics; not all annotations shown for better readability.}
    \includegraphics[width = \textwidth]{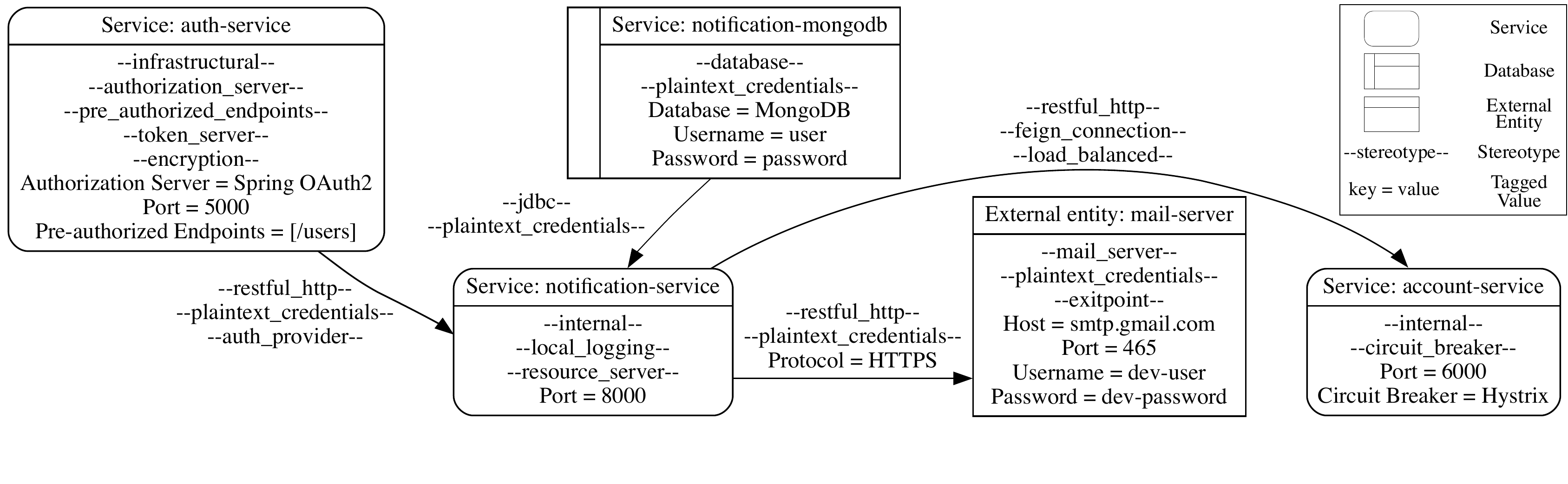}
    \label{fig:model_elements}
\end{figure*}

\noindent
Since DFDs do not have a precise definition as to what model items can be included in them, some variations are used throughout the related literature.
What all DFDs have in common are four core item groups, \textit{processes}, \textit{external entities}, \textit{data flows}, and \textit{data stores}, as already described early in the emergence of this model type~\cite{DeMarco79, Bruza93, Larsen94}.
Additionally, many approaches that work with DFDs add further information to them.
As Sion et al. described~\cite{Sion20}, DFDs comprising only the four basic groups of model items lack expressivity in terms of security concepts and should thus be extended.
As possible solution, multiple approaches add trust boundaries or different kinds of annotations to DFDs~\cite{Berger16, Tuma19, Sion18b, Tuma18} to enhance their usability for security analysis.
In our approach, we add annotations capturing information about implemented security and other features, thereby directly addressing some of the concerns raised by Sion et al.~\cite{Sion20} concerning DFDs for security analysis.

In this paper, DFDs consist of the four base item groups and the additional group of annotations (which are described later in Tables~\ref{tbl:features_nodes} and \ref{tbl:features_flows_external}).
Figure \ref{fig:model_elements} shows an excerpt of the DFD extracted from the application PiggyMetrics\footnote{https://github.com/sqshq/piggymetrics}.
A DFD is a directed graph that contains processes (e.g., microservices), external entities (e.g., clients or third-party systems), data stores (e.g., databases) and information flows among them.
Later on, we collectively refer to services, external entities and data stores as \textit{nodes} in the graph.
In the case of microservice applications, processes (at their finest granularity) correspond to services.
Data stores are also services and, hence, are represented as processes that carry the `database' annotation and have a dedicated visualization.
Finally, note that each element in the diagram can be annotated with additional (security) information.

There are no definitive rules for mapping code elements to a dataflow diagram and in the following paragraphs we describe where the items of each group (services, external entities, information flows and annotations) are discovered in the code.

\emph{Services.}
Microservices are directly mapped to processes (in DFD terminology) and are represented in the figure as rectangles with rounded corners.
In the code, we find services in build files (like Maven) or in Infrastructure as Code files (like  Docker and Docker Compose). 
To differentiate the business logic implemented by the application from supporting, off-the-shelf components, we further distinguish between \textit{internal microservices} (processes), \textit{infrastructural microservices} (processes), and \textit{databases} (data stores). 
Internal services (e.g., `notification-service' and `account-service' in Figure \ref{fig:model_elements}) are developed in the project repository and contain the application's functional code. 
Infrastructural services (e.g., `auth-service' in Figure \ref{fig:model_elements}) mainly use existing code libraries and make only marginal adjustments in order to fit to the application's needs.
Examples include, but are not limited to, message brokers, service discovery solutions, API-gateways, web servers, or configuration servers. 
Databases are services with the additional specification of being used as data stores.
In the DFD, they appear as rectangles with an additional vertical line on the left (see `notification-mongodb' in Figure \ref{fig:model_elements}).

\emph{External Entities.}
These nodes are either clients (i.e., users) or software components that are not deployed with the application, like the Google mail server, a web site, or a GitHub repository to fetch configuration data.
As shown in Figure \ref{fig:model_elements} (`mail-server'), external entities are represented as rectangles.
A client is added to the DFD implicitly to represent the human user when either a gateway or web application is identified.
External software components are called from the application code but there is no deployment descriptor specifying them (e.g., a Dockerfile).
This also includes the case of references to (not co-deployed) databases and results in the creation of an external entity (with the stereotype `database') rather than a data store.

\emph{Information Flows.}
These represent connections between two instances of services and / or external entities over which information is shared.
They are depicted as directed edges in the DFD.
From the perspective of mapping the code to the model, we consider API calls, connections to databases, and message exchanges via message brokers.
Flows are also induced by identifying infrastructure services.
For instance, when a configuration service is found, a flow is created to services that specify it as their source of configuration data.
Information flows have a direction indicated by an arrow tip. 
We consider the direction to be defined by which service or external entity provides data that is used by the other.

\emph{Annotations.}
The DFD elements described above are enriched with additional annotations.
Annotations are \textit{stereotypes} and \textit{tagged values} (key/value-pairs).
In Figure \ref{fig:model_elements}, the annotations are shown as text labels: stereotypes as \textit{--stereotype--}, tagged values as \textit{key = value}. 
Stereotypes represent the semantic purpose of a model element (e.g., a process is an `authorization\_server') or provide some non-functional property of the item (e.g., an information flow carries `plaintext\_credentials').
A list of available stereotypes is presented later in Tables \ref{tbl:features_nodes} and \ref{tbl:features_flows_external}.
Tagged values are used to represent additional attributes, such as ports, API-endpoints, or specific technology solutions.
There are no formal constraints to their content and they are used to provide additional information for a stereotype.
For instance, in the external entity of Figure  \ref{fig:model_elements}, the tags are used to track the username and password used in a plaintext credential (the latter being a stereotype).
The detection of stereotypes and tagged values is done in all types of files throughout the code. 

\emph{Traceability information.}
With our textual analysis approach, it is also possible to easily extract traceability information for all model items described above.
The detection of keywords inherently produces this information as a side-effect and only has to be collected in a structured and coherent way. 
Later in Section \ref{sec:dataset}, we provide an example of how the traceability looks like in our dataset and in the models extracted by our technique.
In summary, the traceability information contains the location of the evidence related to a model item, and is given as the containing file's path and the lines of code.

With these descriptions of the different items in place, we can model DFDs.
Similarly to the standard definition of graphs as a set of vertices and a set of edges, we define a dataflow diagram as a tuple $d = (N, I)$ where $N$ is a set of nodes and $I$ a set of information flows.
A node $n \in N$ consists of a name, type $type \in \{service,\ database,\ external\ entity\}$, set of stereotypes $S$, and set of tagged values $T$.
An information flow $i \in I$ consists of a sender node $sender \in N$, receiver node $receiver \in N$, set of stereotypes $S$, and set of tagged values $T$.
\begin{align*}
    n = (name,\ type,\ S,\ T) \in N \\
    i = (sender,\ receiver,\ S,\ T) \in I
\end{align*}
The set of all stereotypes as defined by our implementation is presented in Tables \ref{tbl:features_nodes} and \ref{tbl:features_flows_external}.

\section{Approach in Detail}
\label{sec:approach}

\noindent As DFDs are representations of software systems, their manual creation from code usually consists of sifting through the codebase for artefacts that indicate the existence of an item in the model. 
Commands (e.g., program instructions, script commands) carry a meaning for the creation of DFDs because of the functionality they invoke in the system. 
We leverage this circumstance to define a technique that extracts DFDs automatically and via a static analysis of the source code.
As will be shown in the following, some items can be detected by looking for single keywords in the code.
For other items, however, the evidence is scattered across the codebase.
For example, detecting a RESTful API connection between microservices requires detecting the creation of the endpoint at the receiving service, the detection of the request at the sending service, and possibly the detection of further properties that define the connection, such as the use of a load balancer, a circuit breaker, or encryption.
Each of these consists of multiple commands in possibly different places.
The challenge thus lies in \textit{connecting the dots}, i.e. deriving coherent statements about a system property (or feature) from a complex set of commands in the code.

In our approach, we employ three different techniques to detect evidence for DFD items: \textit{parsing}, \textit{direct keyword search}, and \textit{iterative keyword search}.

\subsection{Parsing}
\label{sec:parsing}
\noindent Parsing is the simplest of the three techniques.
It is used on structured documents to retrieve information stored in a defined format.
Applied to Infrastructure as Code- and build-files, it reveals the list of microservices in an application and thus provides the initial "building blocks" of the DFD.
When Docker or Docker Compose is used in the analysed application, the images used to build a service can further indicate their functionality.
We use this by checking whether the images of services are on a list of well-known and widely-used ones that imply their functionality, such as Apache httpd, Kafka, or RabbitMQ.

A second important source of information where parsing is applied are configuration files. 
They hold many properties such as the port or name of a service as well as connections between services.
The following excerpt shows part of a \texttt{bootstrap.yml} file.
Parsing it yields the service's name (`notification-service', as seen in Figure \ref{fig:model_elements}) as well as an information flow from another service called `config'.
\begin{center}
\begin{tabular}{c}
\begin{lstlisting}[language=Python]
spring:
  application:
    name: notification-service
  cloud:
    config:
      uri: http:// config:8888
\end{lstlisting}
\end{tabular}
\end{center}

\subsection{Direct Keyword Search}
\noindent Other items are detected via direct keyword search, as single commands can often be sufficient evidence for items in the DFD. 
For instance, Java annotations are a powerful tool to implement complex functionality with a single command in Java applications.
They can thus be leveraged to detect this functionality by merely detecting the keyword of the annotation itself. 
An example for such an annotation is 
\begin{center}
\begin{tabular}{c}
\begin{lstlisting}[language=Java, numbers=none]
@EnableAuthorizationServer
\end{lstlisting}
\end{tabular}
\end{center}
which will start up a Spring OAuth2 authorization server\footnote{https://docs.spring.io/spring-security-oauth2-boot/docs/current/reference/html5}.
Detecting this keyword in the code is evidence for the existence of the service `auth-service', as seen in Figure \ref{fig:model_elements}.

For other model items, the direct keyword search is used as a primer to identify places in the code base where further extraction functions can be applied.
The following example illustrates such an occasion for the information flow from \mbox{`notification-service'} to `account-service' in Figure \ref{fig:model_elements}.
The following line of code associated with the `notification-service' creates a Feign client\footnote{https://cloud.spring.io/spring-cloud-netflix/multi/multi\_spring-cloud-feign.html} connection to the `account-service', which is given as parameter:
\begin{center}
\begin{tabular}{c}
\begin{lstlisting}[language=Java, numbers=none]
@FeignClient(name = "account-service")
\end{lstlisting}
\end{tabular}
\end{center}
A search for `@FeignClient' is used to find lines of code that implement a connection between services in this way.
The target service can then be extracted based on the knowledge of the command's syntax 
and the corresponding documentation, thus, has to be consulted when developing extractors.
The keywords used for the final extraction, such as `name' in the example above, are found in numerous other places in the code with differing meaning because they are quite generic.
A search for the unambiguous keyword `@FeignClient' is thus needed to localize the lines of code where the extraction function can be applied without errors. 

These two ways in which a direct keyword search is applied can detect many of the DFD items and are easily adaptable in new extraction implementations by providing the keywords used for searching and possibly the functions that extract the desired parameters from the findings, which often could be a simple regular expression.

\subsection{Iterative Keyword Search}
\noindent The direct keyword detection ceases to work reliably when developer-chosen identifiers are introduced, because the inherent connection between keyword and functionality is broken. 
Although adhering to naming conventions and choosing descriptive identifiers can retain this connection while improving the code's quality~\cite{Butler09}, automatic techniques have to follow the assumption that identifiers are not descriptive enough to base analyses on them. 
Neglecting the variability introduced by this and detecting model items by matching keywords against identifiers would be unreliable.

As a solution for this problem, we adopt an iterative keyword search technique, which bridges the gap between known keywords and developer-specific identifiers. 
This simple, yet effective technique works in three steps: 
\begin{enumerate}
    \item search for known keywords used as seeds indicating the use of a security feature (e.g., class names from a known security library)
    \item extract new keywords from found instances (identifiers of instantiated classes)
    \item search for new keywords (extracted identifiers) in combination with further known keywords (methods of the classes)
\end{enumerate}

In the example in Figure \ref{fig:model_elements}, `auth-service' uses
Spring framework’s \texttt{BCryptPasswordEncoder} to encode a password. 
After importing the module, an instance of the class is created and named `encoder':
\begin{center}
\begin{tabular}{c}
\begin{lstlisting}[language = Java, numbers=none]
private static final BCryptPasswordEncoder encoder = new BCryptPasswordEncoder();
\end{lstlisting}
\end{tabular}
\end{center}
The iterative search finds this line via the initial keyword `BCryptPasswordEncoder' and extracts `encoder' to be used as keyword for the next iteration.
The second iteration leads to the discovery of the line where the encoding of the user's password takes place further down in the code:
\begin{center}
\begin{tabular}{c}
\begin{lstlisting}[language = Java, numbers=none]
String hash = encoder.encode(user.getPwd())
\end{lstlisting}
\end{tabular}
\end{center}
This is the place where the feature is found. However, the approach provides a more complete traceability of code chain related to this feature.

The search is not restricted to a single file, but can also `jump' to other files. 
For this, when the extracted new keyword contains dot notation, the iterative search checks if there is a file in the same namespace that matches the part before the dot. 
In that case, it looks in this file for the keyword specified after the dot.
Another context switch that is checked for, is the use of environment variables. 
If the extracted identifier begins with a \$ and is enclosed with curly braces, the search mechanism checks for the existence of a \texttt{.env}-file to look up the variable.

In our experience, in the context of Java, this three-step process covered all occurrences of developer-introduced identifiers.
More iterations might be necessary for other languages or artifacts, but the concept remains similar. 
We note that the presented technique is an effective solution for code using known libraries, but does not work automatically for functions implemented entirely in the analysed system, since the known keywords (general purpose functions) do not reveal the overall functionality.
However, security functions should not be implemented in such a way.

\subsection{Combining the Techniques to Extract DFDs}
\label{sec:leverage}

\noindent The three presented techniques are all able to detect (pieces of) evidence for DFD model items in source code.
However, the challenge often lies in piecing the pieces together.
To achieve this, we suggest the use of extractors such that the results from multiple independent detection modules are merged and integrated into a single DFD.
These extractors are designed such that they can be executed automatically one after another.

We introduce the concept of a \textit{technology-specific extractor} (or simply \textit{extractor}) as the detection functionality for a single technology solution.
A technology-specific extractor can detect the existence of an item or a small number of items to be added to the DFD by finding the according evidence in the code. 
Each extractor takes as input the latest state of the extracted DFD and accesses the source code via a search functionality that encompasses the three techniques described above.
It returns the input DFD with possibly added items (nodes, information flows, or annotations to already existing nodes and flows). 
Additionally to extracting DFD items, technology-specific extractors can also produce \textit{intermediate features}, which are part of but not complete evidence for items.
These get passed along with the DFD and if later model extractors detect the missing parts of the evidence, the corresponding item is added.
For example, parsers for configuration files are executed first in the extraction process to gather intermediate features that give context to later extracted DFD items.
An example for such a feature is indicated by the entry:

\begin{center}
\begin{tabular}{c}
\begin{lstlisting}[language = Java, numbers=none]
server.ssl.enabled = True
\end{lstlisting}
\end{tabular}
\end{center}
It will have the effect that all information flows connected to the corresponding service will be marked as using HTTPS. 

Finally, we remark that there are multiple possibilities to implement any given item in the model (e.g., communication encryption). 
Therefore, there can be multiple technology-specific extractors for the same model item, each specific to a different technology.
Both Mayer and Weinreich~\cite{Mayer18} and Mosser et al.~\cite{Mosser20} recognized this need to have a multitude of extraction mechanisms for different technologies in an environment as diverse and fast-changing as microservices.
For instance, Mosser et al.'s approach consists of a set of \textit{probes}, which are manually selected by the user to be run on a certain part of the codebase and enhance a \textit{map} of the software system.
With our technology-specific extractors, we follow a similar approach, however, they do not need to be selected and executed manually but are called automatically.
This requires some consideration of their execution order, because extractors might use intermediate features from other extractors in order to unfold their full functionality.

\subsection{Generalization Beyond Java}
\label{sec:generality}

\noindent This work focuses on Java applications using the Spring framework.
The challenges that static code analysis faces in a field as heterogenous as that of microservices have been identified by others already~\cite{Mosser20, Cerny22, Schiewe22}.
We foresee the interest and necessity to extend our analysis also to other programming languages or frameworks in order to make it useful for a larger part of microservice applications.
To evaluate the feasibility of this task, here we identify the modules that are currently Java- and Spring-specific and reason about their generalizability.

Some of the extractors are already independent of the programming language and framework, because they parse information in configuration files and other generic technologies (e.g. Docker).
For other technology-specific extractors that operate at code level and use the keyword search techniques (direct or iterative), we foresee the need to change the set of keywords they use, e.g., to reflect the identifiers used in the third-party security libraries in other languages and frameworks.
For implementations covering other Java frameworks, this constitutes provisioning of new domain knowledge to Code2DFD, a task that can be done fairly easily by consulting the corresponding documentations. 
Allowing the analysis of other languages than Java requires additional changes, since some search techniques depend on the language syntax of Java. 
For instance, to switch context across files, the search technique assumes the use of the dot notation.
However, as we mostly rely on the analysis of the programming code as text, we speculate that dependencies on language-specific elements are not too difficult to adapt.
To this regard, we emphasize that the design of our tool with its use of technology-specific extractors, the three described detection techniques, and the proposal of what items to consider and how to present them is a solid, valuable foundation for similar tools focusing on other languages or frameworks.

In summary, we believe that porting the existing Java- and Spring-based extractors to other frameworks or languages, like Python, Node, or Go, should be a limited effort, given that our reference implementation for Java Spring is now publicly available and can be used for guidance.
However, this is an interesting aspect that we plan on investigating soon.

\section{The Code2DFD Prototype}
\label{sec:prototype}

\noindent To evaluate our approach, we implemented a prototype in Python that is able to analyse microservice applications written in Java. 
The prototype is called \textit{Code2DFD} and is publicly available at \cite{Code2DFD}.
Following the concept of technology-specific extractors, Code2DFD is structured as a framework: a set of general functionality (the three search techniques presented in Section \ref{sec:approach}, visualization, model management, and similar) builds the foundation on top of which a series of extractors can be hooked.
Our prototype is able to detect the model items presented in Section \ref{sec:mapping} for multiple technology solutions.
We have currently implemented 43 extractors (see the repository for the list and code), covering a large portion of the most prominent technology solutions used for developing microservice applications in Java.
Tables \ref{tbl:features_nodes} and \ref{tbl:features_flows_external} give the complete list of stereotypes extracted by the tool, each with a short description.
At the top layer of the prototype, there is a module that orchestrates the execution of the extractors and combines all detected items into the final DFD.

The tool is implemented as a standalone service with a REST API but can also be run in the terminal for more verbose status information during the analysis. 
As input, it expects a handle to the git repository or path to the local directory containing the microservice application.
In the former case, the repository will be cloned locally first, where the search is performed based on \texttt{grep}.
A keyword search via GitHub's Search API has shown to be much slower.
As output, the DFDs are stored in JSON format and graphically as PNG.

\begin{table*}[!ht]
    \small
	\caption{List of stereotypes for services in DFDs.}
	\label{tbl:features_nodes}
    	\begin{tabular}{p{3.8cm}p{9.2cm}p{4cm}}
    	\hline 
    	\rowcolor[HTML]{D5D3D2}\multicolumn{3}{c}{\textbf{Service (Process or Data Store)}} \\
    	    \hline 
    	    \textit{Stereotype} & \textit{Semantic} & \textit{}\\
    	    \midrule
    	 	administration\_server & Central administration server for all instances. &  \\
        	configuration\_server & Server for managing configurations. &  \\
        	database & Service is used as datastore. &  \\
        	gateway & Dedicated entry point for requests into the system. &   \\
    	    infrastructural & Service that is part of the supporting infrastructure of the system. &  \\
            internal & Service implementing the business logic. &  \\
            in\_memory\_authentication & Service performs in-memory authentication. &  \\
            in\_memory\_datastore & Service uses an in-memory datastore. &  \\
            message\_broker & Message broker for service-to-service communication. & \\
            search\_engine & Service used in the presentation of logging data. &  \\
            service\_discovery & Service registry keeping track of running instances. &  \\
            web\_application & Web application as front-end for the system. &   \\
            web\_server & Service is a web server. &  \\ 
            \midrule
            \textit{Security Stereotype} & \textit{Semantic} & \textit{Security Objective}\\
            \midrule
            authentication\_scope\_all & Service enforces authentication for all incoming requests. & authentication \\
            authorization\_server & Central server for managing authorization decisions. & authorization \\
            basic\_authentication & Basic HTTP authentication enabled for this service. & authentication \\
            csrf\_disabled & Use of CSRF tokens to protect against these attacks is disabled. & authentication \\
            circuit\_breaker & Mechanism to prevent cascading failures. & availability \\
            encryption & Encryption mechanisms are employed. & confidentiality  \\
            load\_balancer & Distributes requests to a service between all instances of it. & availability  \\
            local\_logging & Logging locally at a service. & availability, non-repudiation \\
            logging\_server & Central storage server for log messages. & availability, non-repudiation \\
            metrics\_server & Central storage server for metrics. & availability \\
            monitoring\_dashboard & Central visualization of monitoring information of services. & availability, confidentiality \\
            monitoring\_server & Central service collecting monitoring information from services. & availability, confidentiality \\
            plaintext\_credentials & Plaintext password and/or username found in code. & confidentiality\\
            pre\_authorized\_endpoints & Endpoints for which authorization is checked before execution. & authorization \\
            resource\_server & Resource server in the OAuth protocol. & authorization \\
            ssl\_enabled & Process enforces SSL on connections to it. & confidentiality, integrity \\
            token\_server & Service managing access tokens. & authorization \\
            tracing\_server & Central service collecting tracing information from services. & availability \\
    	\bottomrule
    	\end{tabular}
\end{table*}

\begin{table*}[!ht]
    \small
	\caption{List of stereotypes for information flows and external entities in DFDs.}
	\label{tbl:features_flows_external}
	\begin{tabular}{p{3.8cm}p{9.2cm}p{4cm}}
	\hline 
	\rowcolor[HTML]{D5D3D2}\multicolumn{3}{c}{\textbf{Information Flow}} \\
	    \hline
	    \textit{Stereotype} & \textit{Semantic} &\textit{} \\
	    \midrule
    	feign\_connection & Connection realized using Spring's FeignClient. & \\
        jdbc & Database connection using JDBC protocol. & \\
        message\_producer\_kafka & Sending to a Kafka Queue. & \\
        message\_producer\_rabbitmq & Sending to a RabbitMQ exchange. & \\
        message\_consumer\_kafka & Listening to a Kafka Queue. & \\
        message\_consumer\_rabbitmq & Listening to a RabbitMQ exchange. & \\
        restful\_http & Connection uses restful HTTP. &  \\
        \midrule
        \textit{Security Stereotype} & \textit{Semantic} & \textit{Security Objective}\\
        \midrule
        auth\_provider & Authorization and/or authentication information sent on flow. & authorization\\
        authenticated\_request & Flow between internal services on which requests are authenticated. & authentication\\
        circuit\_breaker\_link & Flow guarded by a circuit breaker. & availability\\
        load\_balanced\_link & Load balanced flow. & availability\\
        plaintext\_authentication & Plaintext password and/or username found in code. & authentication\\
        plaintext\_credentials\_link & Flow over which plaintext credentials are passed. & confidentiality\\
    \hline
    \rowcolor[HTML]{D5D3D2}\multicolumn{3}{c}{\textbf{External Entity}} \\
    \hline 
        \textit{Stereotype} & \textit{Semantic} & \textit{}\\
	    \midrule
        external\_database & Database for which the source code is not accessible. & \\
        external\_website & External website. & \\
        github\_repository & GitHub Repository holding configuration information. & \\
        mail\_server & External mail server. & \\
        user & Represents the human user of the application. &  \\
        \midrule
        \textit{Security Stereotype} & \textit{Semantic} & \textit{Security Objective}\\
        \midrule
        entrypoint & External entity from which information enters the system. & authorization\\
        exitpoint & External entity to which information is sent. & confidentiality\\
        logging\_server & Central storage for log messages. & availability, non-repudiation\\
        plaintext\_credentials & Plaintext password and/or username found in code. & confidentiality\\
        tokenstore & Datastore for access tokens. & authorization\\
	\bottomrule
	\end{tabular}
\end{table*}

\section{Evaluation Methodology}
\label{sec:methodology}

\noindent To evaluate our approach, we created a dataset of DFDs in previous work~\cite{Dataset} by manually inspecting 17 open-source microservice applications.
We report on the creation here as well for completeness and since it was done specifically for the purpose of evaluating our approach.
For the evaluation, the dataset has been partitioned into a reference set (7 apps) and validation set (10 apps), which we used to develop and test our approach, respectively.

\subsection{Dataset Creation} 
\label{sec:dataset_creation}

\noindent No dataset of (microservice) application code paired to DFDs could be obtained from open-source resources, related literature, or other sources.
Most publicly available DFDs are of exemplary nature and do not correspond to actual implementations. 
For the ones that depict real systems, the underlying code is usually not accessible.
We thus curated a dataset ourselves.
The applications were selected from three sources: 
\begin{enumerate}
    \item Lists published by Alshuqayran et al.~\cite{Alshuqayran18}, Marquez et al.~\cite{Marquez18}, and Rahman et al.~\cite{Rahman19}
    \item Search on GitHub for \textit{microservice spring}
    \item Referrals by forums, Google search, etc. during implementation of certain technology-specific extractors
\end{enumerate}

\noindent While the third source did not follow a systematic methodology, applications from this category were only used for reference during development to investigate different implementations of the same concept and make the implementation more generic. 
The validation set consist only of applications from the lists found in literature and the most prominent ones on GitHub based on their relevance.
Table \ref{tab:validation_systems} lists the resulting set of applications. In selecting the apps, we applied the following exclusion criteria:
\begin{enumerate}
    \item[EC1] Programming language is not Java
    \item[EC2] Natural language of documentation is not English
    \item[EC3] System is a framework instead of a standalone application
    \item[EC4] Application uses a technology that is not used by any other application 
    \item[EC5] Application is too big or complex to be manually inspected with the resources at hand in the context of this paper
\end{enumerate}

\noindent We note the limitations these criteria introduce. 
The last two criteria are clear cases of convenience sampling and represent a threat to validity.
We introduce them nonetheless, because of the human effort needed to manually create DFDs for the applications to serve as groundtruth in our evaluation. 
Such effort would be infeasible for larger applications in this context.
However, we note that the exclusion criterion related to the technology (EC4) only applied to two applications (one focusing on deployment with Travis, the other on integration of a Twitter API). 
An example for an application excluded because of EC5 is \textit{Train Ticket}\footnote{https://github.com/FudanSELab/train-ticket}, an application that consists of more than 40 microservices plus databases and over 300.000 lines of code.
The time needed to analyse this application with the level of detail we include in our DFDs is vastly disproportional to the added value of having it in the dataset.
Nonetheless, we hope to include a small number of such big applications in the dataset in the future to observe the prototype's performance. 
We do not expect an observable difference in the number of correctly and incorrectly extracted model items that is related to the size of the application.
Thus, the criteria are seen as acceptable bias.

The DFDs were created manually by reading through the code and documentation.
Prior to the analysis, we created a list of conceptual mappings between source code artifacts and an initial set of DFD items.
These initial DFD items were inspired by the architectural rules proposed by Tukaram et al.~\cite{Tukaram22}.
Each code snippet indicating an item to be added to the DFD was turned into its corresponding model item.
During code inspection, we extended the stereotypes and mappings as we encountered properties that we judged to be of interest for assessing the applications (from a security perspective) until we did not find any new ones anymore.
By re-iterating the inspection of already created DFDs, we achieved consistency across the dataset.
The inspection was performed by the first author, resulting in the extraction of 1989 model items\footnote{We remind the reader that model items include (i) nodes (identified by their name), (ii) information flows (identified by the sender/receiver-pair), (iii) stereotypes, and (iv) tagged values (identified by key/value-pair)}.
We selected a subset of 200 model items following proportional stratified random
sampling, corresponding to about 10\% of the whole dataset.
Each element in the sampled subset has been checked by three further researchers independently.

\subsection{Dataset}
\label{sec:dataset}

\begin{table*}
    \small
    %\centering
    \caption{Applications used for validation with information on size and popularity.
            \\(S = Services, E = External Entities, I = Information Flows, A = Annotations, Tot. = Total Items)}
    \label{tab:validation_systems}
    \begin{tabular}{P{0.2cm}p{4.1cm}P{0.3cm}P{0.3cm}P{0.3cm}P{0.3cm}P{0.4cm}|P{0.8cm}p{6cm}|P{0.6cm}P{0.6cm}}
        \hline
        \rowcolor[HTML]{D5D3D2}\multicolumn{11}{c}{\textbf{Reference set}}\\
        \hline
      
        \textit{ID} & \textit{Application (GitHub handle)} & \textit{S} & \textit{E} & \textit{I} & \textit{A} & \textit{Tot.} & \textit{LoC} & \makecell[c]{\textit{Technologies}} & \textit{Stars} & \textit{Forks}\\
        
        \midrule
        1 & apssouza22/java-microservice & 13 & 2 & 34 & 100 & 149 & 5283 & Docker, Docker Compose, Eureka, Hystrix, Kafka, Logstash, Maven, Nginx, Spring Admin, Spring Config, Spring OAuth, ZooKeeper & 299 & 220 \\
        2 & callistaenterprise/blog-microservices & 15 & 2 & 32 & 126 & 175 & 2786 & Docker Compose, Elasticsearch, Eureka, Gradle, Hystrix, Kibana, Logstash, RabbitMQ, Ribbon, Spring Config, Spring OAuth, Turbine, Zipkin, Zuul & 399 & 308 \\
        3 & fernandoabcampos/spring-netflix-oss-microservices/ & 9 & 2 & 24 & 75 & 110 & 1644 & Docker, Docker Compose, Eureka, Hystrix, Maven, RabbitMQ, Spring Config, Turbine, Zuul & 10 & 11\\
        4 & georgwittberger/apache-spring-boot-microservice-example & 4 & 1 & 6 & 18 & 29 & 604 & Apache httpd, Maven & 7 & 9\\
        5 & mudigal-technologies/microservices-sample & 14 & 1 & 34 & 115 & 164 & 14527 & Consul, Docker, Docker Compose, Elasticsearch, Kibana, Logstash, Nginx, RabbitMQ, \mbox{Weave Scope}, Zuul & 294 & 311 \\ 
        6 & spring-petclinic/spring-petclinic-microservices & 10 & 2 & 28 & 89 & 129 & 3990 & Docker, Docker Compose, Eureka, Grafana, Hystrix, Maven, Prometheus, Ribbon, Spring Admin, Spring Config, Spring Gateway, Zipkin, & 1195 & 1543 \\ 
        7 & sqshq/piggymetrics & 14 & 3 & 37 & 192 & 246 & 9977 & Docker, Docker Compose, Eureka, Hystrix, Maven, \mbox{RabbitMQ}, Ribbon, Spring Config, \mbox{Spring OAuth}, Turbine, Zuul & 11862 & 5697 \\ 
        \midrule
        & \textit{Average reference set} & 11 & 2 & 28 & 102 & 143 & 5544 &  & 1995 & 1157\\
       
        \midrule 
        \hline
        \rowcolor[HTML]{D5D3D2}\multicolumn{11}{c}{\textbf{Validation set}}\\
        \hline
        
        \textit{ID} & \textit{Application (GitHub handle)} & \textit{S} & \textit{E} & \textit{I} & \textit{A} & \textit{Tot.} & \textit{LoC} & \makecell[c]{\textit{Technologies}} & \textit{Stars} & \textit{Forks}\\
        
        \midrule
        8 & anilallewar/microservices-basics-spring-boot & 10 & 2 & 29 & 108 & 149 & 4245 & Docker, Docker Compose, Eureka, Gradle, Hystrix, Ribbon, Spring Config, Spring OAuth, Turbine, Zipkin, Zuul & 645 & 410 \\
        9 & ewolff/microservice & 6 & 1 & 13 & 46 & 66 & 3117 & Eureka, Hystrix, Maven, Ribbon, Turbine, Zuul & 651 & 328\\
        10 & ewolff/microservice-kafka & 7 & 1 & 12 & 56 & 76 & 2979 & Apache httpd, Docker, Docker Compose, Kafka, ZooKeeper & 510 & 272\\ 
        11 & jferrater/tap-and-eat-microservices & 8 & 1 & 16 & 63 & 88 & 1484 & Docker, Docker Compose, Eureka, Hystrix, Maven, Spring Config & 6 & 4 \\
        12 & koushikkothagal/spring-boot-microservices-workshop & 4 & 1 & 6 & 26 & 37 & 638 & Eureka, Maven & 628 & 1012 \\
        13 & mdeket/spring-cloud-movie-recommendation & 6 & 5 & 18 & 93 & 122 & 1800 & Eureka, Hystrix, Maven, Ribbon, Spring Config, Zipkin, Zuul & 13 & 11 \\
        14 & piomin/sample-spring-oauth2-microservices & 5 & 3 & 13 & 78 & 99 & 1028 & Eureka, Ribbon, Spring OAuth, Zuul & 119 & 139 \\
        15 & shabbirdwd53/springboot-microservice & 7 & 2 & 18 & 65 & 92 & 879 & Eureka, Hystrix, Maven, Spring Config, Spring Gateway, Zipkin & 243 & 521 \\
        16 & rohitghatol/spring-boot-microservices & 8 & 3 & 26 & 100 & 137 & 2328 & Docker Compose, Eureka, Gradle, Hystrix, Ribbon, Spring Config, Spring OAuth, Zuul & 1667 & 906 \\
        17 & yidongnan/spring-cloud-netflix-example & 9 & 1 & 30 & 81 & 121 & 1182 & Docker, Docker Compose, Eureka, Gradle, Hystrix, RabbitMQ, Ribbon, Spring Admin, Spring Config, Turbine, Zipkin, Zuul & 782 & 371 \\
        
        \midrule
        & \textit{Average validation set} & 7 & 2 & 18 & 72 & 99 & 1968 &  & 526 & 397 \\
        \bottomrule
    \end{tabular}
\end{table*}

\noindent The dataset consists of 17 DFDs of open-source applications on GitHub. 
Each DFD model includes the application's structure enriched with extensive (security and other) annotations and provides full traceability from model to code, i.e. each item's evidence in the code is tracked by means of its file and line number. 
The models are provided in JSON, PNG, and CodeableModels format. 
The traceability for each model item is presented in a JSON file for each model.

\begin{centering}
\begin{tabular}{c}
\begin{lstlisting}[language=Python,caption={Example for node in JSON representation.}, label={lst:node_example}]
{
    "name": "notification_service",
    "stereotypes": [
        "internal",
        "local_logging",
        "resource_server"
    ],
    "tagged_values": {
        "Port": 8000
    }
}
\end{lstlisting}
\end{tabular}
\vspace{2mm}
\end{centering}

\noindent Listing \ref{lst:node_example} presents an example excerpt of the JSON representation.
The excerpt shows a node in the DFD (corresponding to a node in Figure \ref{fig:model_elements}).

\begin{centering}
\begin{tabular}{c}
\begin{lstlisting}[language=Python,caption={Example for node traceability. URL omitted for better readability.}, label={lst:traceability}]
"notification-service": {
            "file": "*URL*",
            "line": 3,
            "span": "(10:30)",
            "sub_items": {
                "Port": {
                    "file": "*URL*",
                    "line": 13,
                    "span": "(8:12)"
                },
\end{lstlisting}
\end{tabular}
\vspace{2mm}
\end{centering}

\noindent Listing \ref{lst:traceability} shows an example of the traceability information.
The excerpt points to the place in the source code that proves the existence of the model item shown in Listing~\ref{lst:node_example}.

Table \ref{tab:validation_systems} lists the information about the applications in the dataset.
The lines of code (LoC) show a good coverage of small- to medium-sized applications.
The number of services, external entities, information flows, and annotations indicate the size of the corresponding DFD and have a good variation in size and density of annotations.
The list of technologies gives an indication of the extraction complexity, i.e. how many technology-specific extractors are needed for the analysis.
Finally, the numbers of stars and forks on GitHub are given to indicate the applications' popularity.
Although these numbers do not give information about the code's quality, they do suggest that the majority of the applications in the dataset have a decent up to very good acceptance in the community.
The parameters suggest that our dataset is a good and diverse representation of popular small- to medium-sized microservice applications as found publicly on GitHub.

\subsection{Experimental Validation Procedure} 
\label{sec:validation}

\noindent The technique has been evaluated by executing the final prototype on each application in the dataset and analysing the results.
We remark that the authors had access to seven applications in the dataset while developing and tuning the Code2DFD prototype.
In Table \ref{tab:validation_systems}, these applications are listed under the \textit{Reference Set} section (applications 1 -- 7).
The other 10 applications listed under the \textit{Validation Set} section (applications 8 -- 17) have not been analyzed until after the completion of the prototype.
For completeness of information, we report the results for all 17 applications but only consider the validation set when answering the research questions.

We compare the automatically extracted DFDs to the manually curated ones and quantify the approach's performance by counting the items that are correctly and incorrectly identified by the tool (true positives TP and false positives FP), as well as those that go unnoticed (false negatives FN).
Accordingly, we report the precision and the recall achieved by the tool for each individual application, as well as an aggregate over the reference set, the validation sets, and the whole dataset.
We also report precision and recall in two sub-categories: (i) for what concerns only the structural core elements of a DFD and (ii) the security annotations (see Tables~\ref{tbl:features_nodes} and \ref{tbl:features_flows_external}).
Although a key feature of our approach is the extraction of the security-enriching annotations, the technique could also be used outside of security, e.g., by only extracting the core DFD (i.e., the structure of a system consisting of all nodes and connections between them) for program comprehension or security assessment techniques.
Incidentally, this is the reason why we have formulated RQ2.
The category of core items without annotations thus comprises services, external entities, and information flows.
The second category (security annotations) is of particular interest for this paper as this contribution is completely novel (and challenging) with respect to the related work.
For this, we formulated RQ3.

\section{Experimental Results}
\label{sec:results}

\subsection{Quality of the Extracted DFD Models}

\begin{table*}
    \centering
    \caption{Raw evaluation results. S = Services, E = External Entities, I = Information Flows, A = Annotations}
    \label{tbl:numeric_results}
    \begin{tabular}{P{1.7cm}P{0.6cm}P{0.6cm}P{0.6cm}P{0.6cm}P{0.6cm}P{0.6cm}P{0.6cm}P{0.6cm}P{0.6cm}P{0.6cm}P{0.6cm}P{0.6cm}|P{0.85cm}P{0.85cm}P{0.85cm}}
        \toprule
         & \multicolumn{3}{c}{\textbf{S}} & \multicolumn{3}{c}{\textbf{E}} & \multicolumn{3}{c}{\textbf{I}} & \multicolumn{3}{c|}{\textbf{A}} & \multicolumn{3}{c}{\textbf{Overall}}\\
        \textbf{App} & TP & FP & FN & TP & FP & FN & TP & FP & FN & TP & FP & FN & TP & FP & FN \\
        \midrule
        1 & 13 &  &  & 1 & 1 & 1 & 22 & 17 & 12 & 71 & 27 & 29 & 107 & 45 & 42 \\
        2 & 15 & 1 &  & 1 &  & 1 & 19 &  & 13 & 89 & 4 & 37 & 124 & 5 & 51 \\
        3 & 9 &  &  & 2 &  &  & 21 & 2 & 3 & 61 & 2 & 14 & 93 & 4 & 17 \\
        4 & 4 &  &  & 1 &  &  & 5 &  & 1 & 17 &  & 1 & 27 &  & 2 \\
        5 & 14 &  &  & 1 &  &  & 34 &  &  & 107 & 6 & 8 & 156 & 6 & 8 \\
        6 & 10 &  &  & 2 &  &  & 27 & 1 & 1 & 80 & 3 & 9 & 119 & 4 & 10 \\
        7 & 14 &  &  & 3 &  &  & 34 &  & 3 & 173 & & 19 & 224 &  & 22 \\

        \midrule 
        \textbf{Sum 1--7} & 79 & 1 & 0 & 11 & 1 & 2 & 162 & 20 & 33 & 598 & 42 & 117 & 850 & 64 & 152 \\
        \midrule 
        8 & 10 & 1 &  & 2 & 1 &  & 18 & 1 & 11 & 67 & 20 & 41 & 97 & 23 & 52 \\
        9 & 6 &  &  & 1 &  &  & 12 & 1 & 1 & 43 & 1 & 3 & 62 & 2 & 4 \\
        10 & 7 &  &  & 1 &  &  & 12 & 3 &  & 52 & 4 & 4 & 72 & 7 & 4 \\
        11 & 8 &  &  & 1 &  &  & 16 &  &  & 59 & 3 & 4 & 84 & 3 & 4 \\
        12 & 4 &  &  &  &  & 1 & 5 &  & 1 & 23 &  & 3 & 32 &  & 5 \\
        13 & 6 &  &  & 4 &  & 1 & 17 &  & 1 & 84 & 6 & 9 & 111 & 6 & 11 \\
        14 & 5 &  &  & 3 &  &  & 12 & 1 & 1 & 62 & 5 & 16 & 82 & 6 & 17 \\
        15 & 7 &  &  & 2 &  &  & 16 &  & 2 & 57 &  & 8 & 82 &  & 10 \\
        16 & 8 &  &  & 3 & 1 &  & 18 & 1 & 8 & 75 & 12 & 25 & 104 & 14 & 33 \\
        17 & 9 &  &  & 1 &  &  & 26 &  & 4 & 74 & 2 & 7 & 110 & 2 & 11 \\
        \midrule 
        \textbf{Sum 8--17} & 70 & 1 & 0 & 18 & 2 & 2 & 152 & 7 & 29 & 596 & 53 & 120 & 836 & 63 & 151 \\
        \midrule
        \textbf{Sum all} & 149 & 2 & 0 & 29 & 3 & 4 & 314 & 27 & 62 & 1194 & 95 & 237 & 1686 & 127 & 303 \\
        \bottomrule
    \end{tabular}
\end{table*}

\noindent We first look at the approach's performance in terms of correctly extracting DFD items.
Table \ref{tbl:numeric_results} shows the raw results produced by the prototype for the 17 applications.
As a first observation, the detection seems to have worked well in all cases.
All applications show high TP numbers compared to the FN for all groups, meaning that no grave mistakes seem to have occurred.
For the false positives (FP), most cells in the table even have no entries at all. 
Only for the annotations can we see some higher numbers (27 for application 1, 20 for application 8).
For some applications (4, 7, 12, and 15), no FP are detected at all.
We see that only five incorrect nodes (services + external entities) are detected, while the FP for information flows and annotations are higher. 
This was expected, given the challenging nature of these model items.
\begin{table*}
    \centering
	\caption{Precision (P) and recall (R) for each item group. S = Services, E = External Entities, I = Information Flows, A = Annotations, Core Items = S + E + I}
	\label{tbl:pr_results}
    \begin{tabular}{P{1.5cm}P{0.6cm}P{0.6cm}P{0.6cm}P{0.6cm}P{0.6cm}P{0.6cm}P{0.6cm}P{0.6cm}|P{0.9cm}P{0.9cm}|P{0.98cm}P{0.9cm}|P{0.9cm}P{0.9cm}}
        \toprule
         & \multicolumn{2}{c}{\textbf{S}} & \multicolumn{2}{c}{\textbf{E}} & \multicolumn{2}{c}{\textbf{I}} & \multicolumn{2}{c|}{\textbf{A}} & \multicolumn{2}{c|}{\textbf{Overall}} & \multicolumn{2}{c|}{\textbf{Core Items}} & \multicolumn{2}{c}{\textbf{Security Annot.}} \\
        \textbf{App} & P & R & P & R & P & R & P & R & P & R & P & R & P & R \\
        \midrule
        1 & 1 & 1 & 0.5 & 0.5 & 0.56 & 0.65 & 0.72 & 0.71 & 0.7 & 0.72 & 0.67 & 0.73 & 0.78 & 0.47 \\
        2 & 0.94 & 1 & 1 & 0.5 & 1 & 0.59 & 0.96 & 0.71 & 0.96 & 0.71 & 0.97 & 0.71 & 1 & 0.73 \\
        3 & 1 & 1 & 1 & 1 & 0.91 & 0.88 & 0.97 & 0.81 & 0.96 & 0.85 & 0.94 & 0.91 & 1 & 0.67 \\
        4 & 1 & 1 & 1 & 1 & 1 & 0.83 & 1 & 0.94 & 1 & 0.93 & 1 & 0.91 & 1 & 1 \\
        5 & 1 & 1 & 1 & 1 & 1 & 1 & 0.95 & 0.93 & 0.96 & 0.95 & 1 & 1 & 1 & 1 \\
        6 & 1 & 1 & 1 & 1 & 0.96 & 0.96 & 0.96 & 0.9 & 0.97 & 0.92 & 0.98 & 0.98 & 0.88 & 0.93 \\
        7 & 1 & 1 & 1 & 1 & 1 & 0.92 & 1 & 0.9 & 1 & 0.91 & 1 & 0.94 & 1 & 0.92 \\
        \midrule 
        \textbf{Avg. 1--7} & 0.99 & 1 & 0.92 & 0.85 & 0.89 & 0.83 & 0.93 & 0.84 & \textbf{0.93} & \textbf{0.85} & 0.92 & 0.88 & 0.97 & 0.83 \\
        \midrule 
        8 & 0.91 & 1 & 0.67 & 1 & 0.95 & 0.62 & 0.77 & 0.62 & 0.81 & 0.65 & 0.91 & 0.73 & 1 & 0.88 \\
        9 & 1 & 1 & 1 & 1 & 0.92 & 0.92 & 0.98 & 0.93 & 0.97 & 0.94 & 0.95 & 0.95 & 1 & 0.89 \\
        10 & 1 & 1 & 1 & 1 & 0.8 & 1 & 0.93 & 0.93 & 0.91 & 0.95 & 0.87 & 1 & 1 & 0.89 \\
        11 & 1 & 1 & 1 & 1 & 1 & 1 & 0.95 & 0.94 & 0.97 & 0.95 & 1 & 1 & 1 & 1 \\
        12 & 1 & 1 &  &  & 1 & 0.83 & 1 & 0.88 & 1 & 0.86 & 1 & 0.82 & 0.91 & 0.87 \\
        13 & 1 & 1 & 1 & 0.8 & 1 & 0.94 & 0.93 & 0.9 & 0.95 & 0.91 & 1 & 0.93 & 0.84 & 0.72 \\
        14 & 1 & 1 & 1 & 1 & 0.92 & 0.92 & 0.93 & 0.79 & 0.93 & 0.83 & 0.95 & 0.95 & 1 & 0.73 \\
        15 & 1 & 1 & 1 & 1 & 1 & 0.89 & 1 & 0.88 & 1 & 0.89 & 1 & 0.93 & 0.83 & 0.61 \\
        16 & 1 & 1 & 0.75 & 1 & 0.95 & 0.69 & 0.86 & 0.75 & 0.88 & 0.76 & 0.94 & 0.78 & 0.93 & 1 \\
        17 & 1 & 1 & 1 & 1 & 1 & 0.87 & 0.97 & 0.91 & 0.98 & 0.91 & 1 & 0.9 & 0.93 & 0.79 \\
        \midrule 
        \textbf{Avg. 8--17} & 0.99 & 1 & 0.9 & 0.9 & 0.96 & 0.84 & 0.92 & 0.83 & \textbf{0.93} & \textbf{0.85} & 0.96 & 0.89 & 0.88 & 0.75 \\
        \midrule
        \textbf{Avg. all} & 0.99 & 1 & 0.91 & 0.88 & 0.92 & 0.84 & 0.93 & 0.83 & \textbf{0.93} & \textbf{0.85} & 0.94 & 0.88 & 0.93 & 0.79 \\
        \bottomrule
    \end{tabular}
\end{table*}

Table \ref{tbl:pr_results} shows the computed precision and recall, allowing us to put the raw numbers into perspective.
A first observation is a very good result with an overall average precision of 93\% and recall of 85\%.
As mentioned before, we observe a high precision and recall for the nodes but slightly lower numbers for the information flows and, particularly, the annotations.
This is explicitly evident in the values for the extraction of core items without annotations, with a precision of 94\% and recall of 88\% .
With 79\%, the security annotations show a lower recall than the other groups, where 47\% is the worst observed recall (application 1).
Looking at the results for individual applications, we notice that some achieve an overall precision or recall of only around 70\%. 
The lowest overall value is a 65\% recall for application 8.

\noindent\fbox{
    \parbox{0.97\linewidth}{
    Table \ref{tbl:pr_results} (Applications 8 -- 17) allows us to provide the following answers to RQ1, RQ2, and RQ3:
    \begin{description}
        \item [RQ1] We observe a precision of 0.93 and recall of 0.85 for the extraction of all DFD items.
        \item [RQ2] We observe a precision of 0.96 and recall of 0.89 for the extraction of DFD core items.
        \item [RQ3] We observe a precision of 0.88 and recall of 0.75 for the extraction of the security-relevant annotations.
    \end{description}
    }
}

In summary, the results show that our approach is a promising solution for reliably extracting dataflow diagrams.
In Section \ref{sec:discussion}, we elaborate further on how to interpret the results.

\subsection{Time Performance of Model Extraction}
\label{sec:performance}

\begin{figure}
    \centering
    \caption{Execution times on all applications, average over five execution runs.\\
    Lines are min to max, blue markers are average over the five runs.}
    \includegraphics[width = 0.9\linewidth]{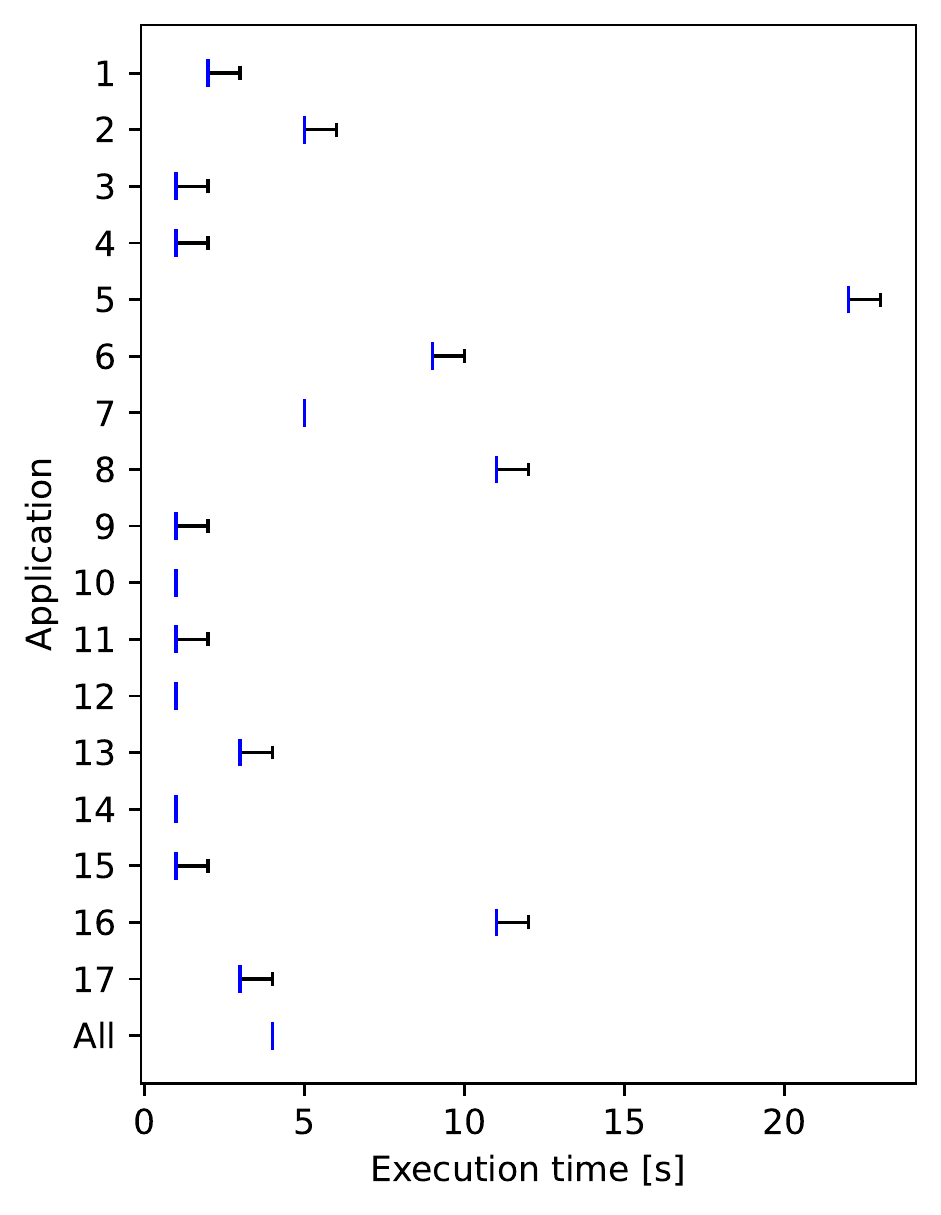}
    \label{fig:execution_time}
\end{figure}

\noindent Since a possible use case of our approach is the adoption in CI/CD pipelines, security delta-evaluation, evaluation of apps, and other time-sensitive scenarios, the prototype's performance concerning its execution time is of interest as well. 
We ran the DFD extraction five times on each application and took the average of the complete execution time, excluding the cloning of repositories from GitHub.
We performed the experiments on a MacBookPro with 2GHz processor and 16GB RAM. 

The results are shown in Figure \ref{fig:execution_time}. 
Per application, the extraction times vary between 1 and 22 second, with an overall average of 4 seconds.
Most apps lie under the 5 seconds mark.
The standard deviations for all applications is under 1 second, indicating a very stable extraction time.
When comparing Figure \ref{fig:execution_time} and Table \ref{tab:validation_systems}, the numbers show no obvious correlation between the execution time and size of the corresponding application or model, number of used technologies, or other parameter.
An analysis of what factors influence the execution time is left open for future work.

\noindent\fbox{
    \parbox{0.97\linewidth}{
    Figure \ref{fig:execution_time} allows us to provide the following answer to RQ4:
    \begin{description}
        \item [RQ4] We observe a maximum model extraction time of 22 seconds, with an overall average of only 4 seconds.
    \end{description}
    }
}

Overall, the execution times illustrate that our (fully automated) approach is applicable in fast-paced development pipelines and other time-sensitive scenarios.

\subsection{Threats to Validity}
\label{sec:threats_to_validity}

\noindent Some limitations in our evaluation stem from the lack of available data and the resulting creation of a manually curated dataset.
Extracting the DFDs and their model elements from the code of the applications has been done manually by one researcher and checked by three others.
Although we are quite confident that the models and the traceability information are correct, bias and material errors are always a possibility.

We are also aware that the dataset is of limited size (17 applications), the applications have been selected by means of convenience sampling, and that a few applications are not particularly complex.
Therefore, the results we report might not entirely transfer to other real-world applications.
In this respect, the lack of data for validation is a well-known issue in the model-based security community.
While some papers do present DFD models extracted from medium-to-large size application, they are not useful in our case as (i) they are not related to microservices and (ii) they do not provide the traceability to code information.

Finally, our validation results might be specific to the technologies that we support in our extraction framework. 
While the stereotypes we consider (as listed in Tables \ref{tbl:features_nodes} and \ref{tbl:features_flows_external}) are meaningful and comprehensive for security, they evidently do not cover all technological and security features one could encounter in microservices written in Java.

\section{Discussion of Results}
\label{sec:discussion}

\noindent In this section, we further elaborate on the causes for the false positives and false negatives encountered in the evaluation as shown in Table \ref{tbl:numeric_results} and talk about the corresponding precision and recall presented in Table \ref{tbl:pr_results}.
We also elaborate on which items these errors apply to in order to better assess what results can be expected when employing our approach. 
We further put the results into context regarding the usability of our approach for security analysis use cases.

\textbf{Nodes.}
The correct detection of all but two nodes is assuring but not surprising, since they are mostly detected by the simplest search technique, parsing.
Alternatively, the nodes are found by references to an external entity.
The group of services does not have any false negatives, but there are four applications for which an external entity is not detected.
Looking at the causes, these failing detections are acceptable for the state of the prototype.
For example, one is an undetected database, for which only an ambiguous line of code and a comment in the \texttt{README} file are evidence.
A failing detection of this item can be regarded as acceptable error or even as out-of-scope of a static analysis approach, definitely out-of-scope of our approach.
Similar findings hold for the false positives.
In two cases, the tool creates two DFD items for the same database because two different names are used for it in the application, in another case a false positive node is a GitHub repository which is detected as an external entity because it is referenced in a configuration file as external configuration repository, but not present in the ground truth because it is the same repository where the complete code lies. 
It is debatable if this should even count as a false positive, but to keep the results conservative we count it as such.

\textbf{Information flows.}
A precision of 92\% and a recall of 84\% for this group shows that the techniques are applicable for complex detections such as that of message broker connections, where outgoing and incoming endpoints have to be detected individually and matched while taking the syntactical rules of queues, exchanges, and routing keys into account.
We see this as affirming result for our approach.

The reasons for some false positives and false negatives lie in the direction of the information flows.
For the ground truth, we decided on the direction based on the source of the data asset that is exchanged.
In some cases, the prototype detects the flow but with the wrong direction, which leads to both a false positive and a false negative. 
This highlights the need to improve the corresponding extractors.
Other false positives happen because the prototype detects connections of services to themselves. 
For example, a monitoring server can have the scope `all services'. 
A flow from that service to itself will be detected in this case.
However, there are plain errors as well, where the detection failed due to inaccurate extractors.
The only case for which the number of false positive information flows is concerningly high is for application 1.
Here, the above mentioned problem of flows being detected with the wrong direction happened multiple times in addition to a faulty implementation of the technology-specific extractor for Apache ZooKeeper.
An assessment of the false positives and false negatives of all applications (especially applications 1, 2, 8, and 16 with their comparatively low recall) yields, that the majority of the errors should be addressable with minor adjustments to the existing extractors or by adding some.
We believe that achieving better results while not overfitting to a dataset and keeping generality is possible with a more sophisticated implementation of our approach.

\textbf{Annotations.}
The annotations are an important group of items to evaluate our approach since they are a distinguishing feature compared to related approaches and many of them carry the security-specific information.
Results of 93\% precision and 83\% recall are thus seen as a very good and promising outcome.
False positives often stem from the reduced scope of implemented mechanisms, which is often hard to grasp with the technology-specific extractors. 
For example, the extractor for circuit breakers marks all outgoing flows from a service it detects as being guarded by the circuit breaker, which is not always correct.
Other false positives are due to legacy code found in some applications, which are detected by the extractors but do not have any effect in the application.
As for the information flows, the majority of undetected annotations is due to the implementation of our prototype, with flaws in the extractors or simply missing extractors for some technology solutions.
We did not encounter any annotations for which we believe a detection with our approach to be impossible.

\textbf{Security annotations.}
To more specifically evaluate the performance concerning the security analysis, we look at the precision and recall for the subset of security-relevant annotations.
Compared to the other cases, the recall is slightly lower (79\%).
The reason lies in the detection complexity of many of the security annotations. 
Especially authentication and authorization are concerns that require detailed configurations at numerous places in the code, which makes the detection of the corresponding stereotypes particularly challenging. 
However, the numbers also show that these more complex annotations can still be detected with our approach, although more work in this area could be needed in the future.

\textbf{Validation vs. reference set.}
As described in \ref{sec:validation}, we used the reference set for testing and tuning during the development of Code2DFD. 
Accordingly, we expected a degree of `over-fitting' in the reference set accompanied by lower performance results in the validation set.
However, the drop in performance did not happen.
One explanation lies in the relatively small dataset, for which the overall results can be swayed by the outcome of single applications. 
Also, a larger dataset might cover applications where the code displays more variations in the way the technologies are used (hence causing more false negatives).
For now, we report the good results even for unseen applications and plan a more extensive evaluation for future work.

\section{Comparison to other approaches}
\label{sec:comparison}

\noindent 
To provide context for our prototype's performance, we compare its results with those of other related approaches.
We identified available and executable tools of the approaches for architecture reconstruction that we present later in Section~\ref{sec:related_work} (listed in Table~\ref{tbl:related_work}).
Two of these provide implementations of their approach that we were able to obtain and execute.
The remaining approaches in Table~\ref{tbl:related_work} either do not present and make available an implementation, or the approach is dynamic and thus not applicable for this comparison.
Moreover, in those publications where a comparable implementation is absent, there are no evaluation results reported that could be presented here as a comparison.
We ran the two identified tools on all 17 microservice applications in our dataset of microservice applications and evaluate the outcomes.
The results reported in this subsection are thus verified by us on the same dataset of applications that we evaluated our own tool on.

The two identified tools do not consider the extraction of (security) annotations.
Therefore, we only compare the architecture reconstruction.
We were unable to find any relevant work that focuses on detection of security features or other comparable system properties.

\begin{table*}
\centering
	\caption{Precision and recall in architecture reconstruction for the related approaches. S = Services, E = External entities, I = Information Flows.}
	\label{tbl:approach_comparison}
    \begin{threeparttable}
        \begin{tabular}{p{5.5cm}|P{2cm}|P{0.8cm}P{0.8cm}P{0.8cm}P{0.8cm}P{0.8cm}P{0.8cm}|P{0.8cm}P{0.8cm}}
            \toprule
             & \textbf{Analyzed} & \multicolumn{2}{c}{\textbf{S}} & \multicolumn{2}{c}{\textbf{E}} &\multicolumn{2}{c|}{\textbf{I}} & \multicolumn{2}{c}{\textbf{Overall}} \\
            \textbf{Approach} & \textbf{Applications} & P & R & P & R & P & R & P & R\\
            \midrule
            Prophet (Bushong et al.~\cite{Bushong21, Bushong22}) & 5 / 17\tnote{1} & 0.59 & 0.22 & 0 & 0 & 0.75\tnote{1} & 0.07\tnote{1} & \textbf{0.61} & \textbf{0.15} \\
            MicroDepGraph (Rahman et al.~\cite{Rahman19}) & 10 & 1 & 0.86 & 1 & 0.13 & 0.98 & 0.51 & \textbf{0.99} & \textbf{0.58} \\
            \midrule
            Code2DFD & 17 & 0.99 & 1 & 0.91 & 0.88 & 0.92 & 0.84 & \textbf{0.94} & \textbf{0.88} \\
            \bottomrule
        \end{tabular}
        
        \begin{tablenotes}
        \small
            \item[1]: The tool produced a list of microservices for 17 apps, but communication diagrams only for five of them. Results for \textbf{I} are only for these five.
        \end{tablenotes}
    \end{threeparttable}
\end{table*}

\subsection{Compared Tools}
\noindent
The first tool, \textit{Prophet}~\cite{Bushong21, Bushong22}, extracts microservices and communication links between them by leveraging enterprise standards, i.e. knowledge about standard components and constructs created by popular development frameworks.
According to the authors, the semantic purpose of relevant components in code can be identified via naming conventions and other metadata, which leads to the detection of microservices.
Communication links between the services are detected by identifying API endpoints and API calls to other services in the same way.
The tool produced complete results for five of the 17 applications. 
For all others, it only extracted the list of services it detected in the applications. 
In the calculation of precision and recall for the information flows, we only consider these five successful extractions to provide a fair comparison.

The second tool, \textit{MicroDepGraph}~\cite{Rahman19}, parses Docker Compose and Java files to detect microservices and communication links between them.
The tool parses a Docker Compose file (if one exists) to extract the list of microservices and communication links (from specified service dependencies).
Additionally, direct API calls between services are extracted from Java source code by checking for Spring Boot annotations that define API endpoints and for requests to these endpoints from other services.
The reliance on the use of Docker Compose led to a limitation of the analysis to 10 of our 17 applications. 
We use these 10 to calculate the results.
In addition, minor changes to the code of the analysed applications were needed, because the tool does not consider all legal syntaxes of Docker Compose and some applications had to be adjusted to be correctly parsed by the tool.

\subsection{Comparison Results}

\noindent
Table~\ref{tbl:approach_comparison} shows the results of Code2DFD and the two compared tools for the extraction of the microservices' architecture (i.e., services, external entities, and information flows; annotations omitted since the related approaches do not consider them).

The first tool achieves an overall precision of 0.61 and recall of 0.15.
The tool produced many false positive services from which the names correspond to directories in the applications' repositories. 
The precision for just the information flows (calculated on five applications for which a communication diagram was successfully created by the tool) is slightly higher, but the recall is 0.07.
Overall, this approach does not offer a reliable architecture extraction for the applications in our dataset.

The second tool produces an overall precision of 0.99 and recall of 0.58.
The tool extracts most information via simple parsing of Docker Compose files.
These files are usually a reliable source for architecture reconstruction, but contain limited information.
This leads to a high precision because false positives rarely happen in this process, but also to a rather low recall.

Compared to our Code2DFD, the second approach achieves a slightly higher precision (0.99 against 0.94). 
However, while we observe a recall of 0.88 in Code2DFD, the second approach achieves a lower recall of 0.58.
The first approach produces significantly lower results than Code2DFD on our dataset.
In conclusion, our approach proves to be competitive with the two compared approaches regarding the architecture reconstruction.
It shows a slightly lower precision than approach 1 and outperforms both approaches significantly concerning the recall.

\section{Related Work}
\label{sec:related_work}

\begin{table*}[!ht]
	\caption{Extraction techniques employed in the related work for architecture reconstruction.}
	\label{tbl:related_work}
    \begin{tabular}{P{0.8cm}|p{4.1cm}|p{4.5cm}|p{7.1cm}}
        \toprule
        ID & Approach & Services & Information Flows \\
        \midrule
        1 & Alshuqayran et al.~\cite{Alshuqayran18} & static parser & inject Zipkin tracing \\ 
        2 & Bushong et al.~\cite{Bushong21, Bushong22} & static & static detection of API endpoints\\
        3 & Granchelli et al.~\cite{Granchelli17a, Granchelli17} & static parser & injected TcpDump, existing service discovery \\ 
        4 & Kleehaus et al.~\cite{Kleehaus18} & existing Consul or Eureka  & Google Dapper \\ 
        5 & Ma et al.~\cite{Ma19} & existing Eureka  & Java annotations \\ 
        6 & Mayer and Weinreich~\cite{Mayer18} & static parser, Swagger & own interceptor modules \\ 
        7 & Rademacher et al.~\cite{Rademacher20} & static parser & Java annotations \\
        8 & Rahman et al.~\cite{Rahman19} & static parser & Spring annotations \\  
        9 & Soldani et al.~\cite{Soldani21} & static parser & Kubernetes network traffic \\
        10 & Walker et al.~\cite{Walker21} & static (source and byte code) & static (source and byte code) \\
        
        \bottomrule
    \end{tabular}
\end{table*}

\noindent 
With its extraction of extensively annotated DFDs that can be used to analyse software systems' security, our approach combines different scientific research fields.
The extraction of the core DFDs (i.e. their nodes and edges) falls under the field of architecture reconstruction, the extraction of annotations relates to feature identification, and the enabling of a security analysis with the DFDs has related work in model-based security.
In the following, a comparison to the literature is presented separately for these three fields.

\textbf{Architecture reconstruction.} Multiple approaches have been presented in literature that employ different combinations of static and dynamic analysis to recover the basic architecture of microservice applications.
Table \ref{tbl:related_work} lists those related to our work and shows the techniques used to extract the services and information flows, thus providing an overview of the differences to our approach. 
Not all of the presented approaches are automated, but they at least claim to be automatable or mention it as future work.
A detailed description is given in the following (mentioned approach IDs correspond to those in the table).

Most approaches retrieve the list of services in the same way we do in Code2DFD, by parsing Docker, Docker Compose, Maven, or Kubernetes manifest files (approaches 1~\cite{Alshuqayran18}, 2~\cite{Bushong21, Bushong22}, 3~\cite{Granchelli17a, Granchelli17}, 6~\cite{Mayer18}, 7~\cite{Rademacher20}, 8~\cite{Rahman19}, and 9~\cite{Soldani21}).
Approach 10~\cite{Walker21} works statically as well, but analyses control-flow graphs and program dependency graphs of the applications.
Approaches 4~\cite{Kleehaus18} and 5~\cite{Ma19} detect microservices via service discovery technologies at run-time.
For this, they query service discovery services to retrieve the list of currently deployed instances of microservices.
The two approaches are limited to implementations using \textit{Consul} (only approach 4) or \textit{Eureka} (both) that already exist in the analysed applications, thus reducing the scope to such applications where this holds true.

In the related work, information flows are either detected statically or dynamically with different tracing implementations.
Approaches 2, 5, 7, and 8 leverage Java annotations to statically detect the flows, following one of the detection techniques we use in Code2DFD.
We note the difference in scope between these approaches and ours. 
Apart from further detecting information flows via methods not connected to annotations (i.e. direct API calls with e.g. \textit{RestTemplate}), we also consider more annotations in the detection.
From the approaches using dynamic analysis to retrieve information flows, approach 3 uses existing service discovery services and an injected monitoring tool.
Approach 1 injects a \textit{Zipkin} tracing server, a technology that is used by some of the applications in our dataset as well, while approach 4 injects \textit{Dapper}, another tracing technology. 
By letting all existing services send tracing data to the injected service, the information flows can be observed.
Approach 6 also injects components to trace information, but their interceptor modules are developed specifically for this purpose. 
Finally, approach 9 specializes on deployment via Kubernetes and observes the network traffic to obtain communication links between services.
All these dynamic approaches have the drawback that they require changes to be made to the existing application that is to be analysed.
Further, they can only detect information flows that get executed during the analysis, as all dynamic approaches do.

It is evident, that there are similarities between the related work and our approach considering the extraction of the structure of microservice applications.
Approaches 2, 7 and 8 work fully statically and follow the same base idea for retrieving services and information flows.
However, apart from the larger scope and lightweight static analysis, another clear distinguishing feature of our work is the extraction of application properties beyond the basic structure (and with focus on security), in order to give a more detailed view of the software system.
Approach 1 is the only other approach that is also detecting load balancers, gateways, service discovery, circuit breakers, and access tokens. These features, however, are a small subset of what we extract, especially from a security standpoint (compare Tables \ref{tbl:features_nodes} and \ref{tbl:features_flows_external}). 
Finally, and most notably, all approaches in Table \ref{tbl:related_work} require human input and, most of them, a heavyweight dynamic analysis orchestration, both of which we avoid.

Concerning the structure of the proposed technique, Mosser et al.~\cite{Mosser20} present an approach that follows a similar architecture as Code2DFD. 
Their \textit{Animaxander} consists of probes that extract a map of the microservice application, showing parallels to our technology-specific extractors.
However, it relies on human input to a large extent and does not focus on security. 

\textbf{Feature identification.}
No related work addressing automated (security) feature identification specifically for microservice applications could be found in the literature.
In fact, Assun\c{c}\~{a}o et al. identified this as a challenge for future work on microservices~\cite{Assuncao20}.
However, there are approaches that consider feature identification in other contexts without a focus on microservices.
Many approaches for the identification and traceability of features and their location in source code have been made in the fields of program comprehension and software maintenance and evolution~\cite{Dit13, Rubin13}.

Many static approaches for feature identification rely on the existence of a source of information that is created by the developers beyond the functional code.
For example, many approaches leverage annotations in the source code that are added specifically for the purpose of tracing and visualizing features~\cite{Abukwaik18, Andam17, Bergel21, Entekhabi19, Martinson21, Seiler17}.
Since the code has to be annotated first (either during implementation or retroactively), these approaches do not solve the problem of automatically detecting the features on unmodified code or code that wasn't developed with subsequent feature identification in mind.

Other approaches omit this dependence on dedicated annotations by leveraging already existing information in code.
They use natural language processing, machine learning, pattern matching, or other techniques to identify features.
For this purpose, identifiers, comments in the code, or documentations are commonly used as input~\cite{Burger18, Eaddy08, Marcus04, Savage10, Zhao06}.
These approaches suffer from a reliance on the developers choosing meaningful identifiers and writing descriptive comments and documentation, otherwise they cease to work.
We avoid this in our approach with the help of the iterative keyword search and codification of domain-knowledge about technologies via the technology-specific extractors.

We conclude, that our approach is novel in regard to identifying (security) features in microservice applications and that with this it addresses one of the challenges for microservice analysis posed by Assun\c{c}\~{a}o et al.~\cite{Assuncao20}.
Further, the presented related approaches rely on information that is chosen by developers and as such not completely reliable, which we avoid.

\textbf{Model-based security.} 
Some approaches exist that use DFDs to perform security analysis at model level.
Abi-Antoun et al.~\cite{Abi-Antoun07} compare a security-annotated DFD extracted from the implementation code of a system against one that represents the design specifications. 
By checking conformance between the two, architectural drift can be detected.
The approach further enables a security analysis by formulating security properties based on the annotations in the DFDs that can be checked.

Berger et al.~\cite{Berger16} introduce eDFDs (short for extended DFDs), an extension of DFDs that also captures properties like trust boundaries and better tracking of data flows. 
The authors collect threats in a knowledge base and use the Microsoft Threat Modeling technique to find architectural weaknesses. 
Tuma et al.~\cite{Tuma19} introduce SecDFDs (short for Security DFDs), an extension to DFDs that includes further information like security policies, assets' importance, and security labels (corresponding to private or public information). 
Based on security contracts defining how security labels change for specific operations, the assets are tracked through the SecDFD and violations of security policies can be detected. 
In both approaches, the user needs to understand the analysed application well, as eDFDs and SecDFDs have to be created manually.

Sion et al.~\cite{Sion18} present an approach in which architectural patterns can be formulated to capture implemented security or privacy countermeasures.
The patterns are expressed by referring to the elements of corresponding DFDs.
When performing a security analysis such as STRIDE with the DFDs as input, the patterns support the analysis by removing threats from consideration that are already mitigated by a security solution.
Our extracted DFDs might be a good fit for such architectural patterns, as their extensive annotations should allow the formulation of rich and expressive patterns.

Instead of extending DFDs with additional information as the approaches above do, Faily et al.~\cite{Faily20} propose to put them into context of other usability and requirement models.
They argue, that the simplicity of basic DFDs should be preserved and that alignment with other models should be strengthened. 
The authors illustrate this idea on the example of taint analysis.

Berger et al.~\cite{Berger13} proposed ArchSec, a solution for automatically extracting the security architecture of Java applications. 
ArchSec only works for J2EE and Android applications, and not for microservices.
Furthermore, the tool expects that the system's decomposition into the top level components (and the segmentation of the codebase into the corresponding components) is provided as input by a human.

In conclusion, to the best of our knowledge no approach has been proposed in the literature that allows a fully automated extraction of dataflow diagrams or similar model representations of microservices which are enriched with additional, extensive (security) annotations and thus allow an assessment of the corresponding application's security.
Finally, we further emphasize the extent of our evaluation with the comparison to other tools and the comparably large dataset, which is unrivalled by the mentioned related approaches.

\section{Conclusion}
\label{sec:conclusion}

\noindent This paper presents an approach for extracting dataflow diagrams enriched with extensive (security and other) annotations from microservices' source code.
The approach analyzes applications statically and interprets the code (source code, configuration files, scripts, and so on) as text. 
The implementation is focused on Java applications and can be applied to systems that use the technologies for which an extraction implementation is provided. 
However, we discussed how the approach can be extended to include other languages and technologies.
In our validation results, we observe a very satisfying performance with a precision above 90\% and recall of 85\%. This outperforms two related approaches in architecture reconstruction that we executed on the same dataset.

In future work, we plan on extending our prototype to support polyglot applications.
Further, we want to increase the size of our dataset with additional applications, as the lack of data is an issue in the model-based security community.

\section*{Acknowledgement}
\noindent The authors thank Catherine Tony, Quang Cuong Bui, and Torge Hinrichs for their support in validating the dataset.

This work was partly funded by the European Union's Horizon 2020 programme under grant agreement No. 952647 (AssureMOSS).

\bibliographystyle{elsarticle-num} 
\bibliography{dfd_extraction}

\end{document}